\documentstyle[twocolumn,prc,aps,epsfig]{revtex}
\input epsf
\newcommand{\be}{\begin{eqnarray}}
\newcommand{\ee}{\end{eqnarray}}
\def\lsim{\mathrel{\rlap{\lower3pt\hbox{\hskip1pt$\sim$}}
     \raise1pt\hbox{$<$}}} %less than or approx. symbol
\def\gsim{\mathrel{\rlap{\lower3pt\hbox{\hskip1pt$\sim$}}
     \raise1pt\hbox{$>$}}} %greater than or approx. symbol

\def\tl{\tilde\lambda}
\def\tc{\tilde{c}}
\def\tv{\tilde{v}}
\def\Fex{\bf{F}_{\rm ex}}
\def\Fdh{\bf{F}_{\rm DH}}
\begin{document}

\twocolumn[\hsize\textwidth\columnwidth\hsize\csname 
@twocolumnfalse\endcsname

\title{Classical Strongly Coupled QGP II:\\Screening and
 Equation of State}

\author{Boris A. Gelman$^{1,2}$, Edward V. Shuryak$^{1}$ and 
Ismail Zahed$^{1}$}
\address {$^{1}$ Department of Physics and Astronomy\\
State University of New York, Stony Brook, NY 11794-3800, \\
$^{2}$ New York City College of Technology, Brooklyn, NY 11201 }

%\date{\today}
\maketitle
\begin{abstract}
We analyze the screening and bulk energy of a classical
and strongly interacting plasma of color charges, a model we
recently introduced for the description of a quark-gluon plasma
at $T=(1-3)T_c$. The partition function is organized around the
Debye-H$\ddot{\rm u}$ckel limit. The linear Debye-H$\ddot{\rm u}$ckel 
limit is corrected by a virial expansion. For the pressure, the 
expansion is badly convergent even in the dilute limit.
The non-linear Debye-H$\ddot{\rm u}$ckel theory is studied numerically
as an alternative for moderately strong plasmas. We use Debye theory 
of solid to extend the analysis to the crystal phase at very strong
coupling. The analytical results for the bulk energy per particle
compare well with the numerical results from molecular dynamics simulation
for all couplings.
\end{abstract}
\vspace{0.1in}
]
\begin{narrowtext}
\newpage

\section{Introduction}

QCD at high temperature is believed to be in a quark gluon 
plasma (QGP) phase, whereby color charges are screened rather than
confined~\cite{Shu_QGP}. Asymptotic freedom guarantees that for
$T\gg \Lambda$ (the QCD cutoff) the QGP is weakly coupled (wQGP)
with dressed quarks and gluons behaving as quasi-particles near
the ideal gas limit.

The QGP has been intensely sought by dedicated  experiments 
at CERN SPS and more recently at BNL RHIC collider. Extensive
analysis of RHIC data have revealed collective effects known as
radial and elliptic flows. Their hydrodynamical explanation
~\cite{hydro}  have suggested that the QGP is a near-perfect 
liquid~\cite{Shu_liquid}, promptly formed in heavy ion collisions 
at temperatures of the order of $T\approx 2\, T_c$. As
all dissipative lengths in the QGP seem to be short, it
is not a weakly coupled gas-like phase but more
like a liquid-like phase.

Recently, two of us~\cite{SZ_newqgp} have suggested that the interaction
of quasiparticles in the relevant temperature range probed at RHIC
is strong enough to form multiply marginal bound states. Some of these
states (charmonium) were recently reported in lattice simulations
\cite{charmonium} at temperatures as high as $T\approx 2\,T_c$. 
Many more colored and uncolored states were predicted,
and remain to be analyzed by lattice simulations. The QGP in this regime 
will be referred to as a strongly coupled QGP (sQGP). First principle 
calculations of the sQGP properties can be done for supersymmetric extensions 
of QCD, $N=4$ SYM, via the AdS/CFT correspondence.

In a recent paper \cite{GSZ_I}, hereby referred to as I, we suggested a 
classical nonrelativistic model to help understand and describe some
relevant features  of the sQGP. Essentially it is the transport properties 
of the sQGP, which are difficult or impossible to access through lattice 
simulations. The model consists of massive nonrelativistic quarks and
gluons, which interact via color Coulomb interactions plus
some repulsion, for stability. The color charges are assumed to be large 
and classical obeying Wong equations of motion. In a way, we are 
suggesting that quantum effects in the QGP are basically reduced to
generate thermal-like masses, cause the effective coupling to run to larger 
values at smaller $T$, and add the ``localization energy'' to
the Coulomb interaction. This model, referred to as classical QGP (cQGP)
was studied in I using molecular dynamics simulations.

In this paper we provide some analytical analysis of the
static bulk properties such as the pressure and the energy density 
of the cQGP thereby unraveling its phase structure at weak,
intermediate and strong coupling, where is is
in a gas, liquid and solid phases respectively. 
Similar approaches in the context of solutes of different natures 
where also considered~\cite{SOLUTES}.

In section II we review briefly the cQGP model. In
section III we carry an exact low density expansion of the cQGP 
around the linearized Debye-H$\ddot{\rm{u}}$ckel limit to sixth-order 
in the particle concentration. In section IV 
we go beyond the linear Debye-H$\ddot{\rm{u}}$ckel theory and study screening
in a nonlinear regime. In section V we study another version
of screening, via the crystal phase. In section VI
we use Debye theory 
of solids to extend the Debye-H$\ddot{\rm{u}}$ckel result for dilute but
screened gases all the way to the crystalline phase. Our
analytical results for the excess energy and pressure compare favorably to the
results from molecular dynamics at all couplings. Our conclusions
are summarized in section VII.

\section{Classical QGP Model}

For quarks and gluons with large thermal energies of the
order of $\pi T$, we may consider the cQGP
as a system of non-relativistic charges interacting
solely through longitudinal color electric fields. The magnetic
effects are subleading in the non-relativistic limit.
Specifically, the Hamiltonian for the cQGP reads

\be
H=\sum_{\alpha\,i}\frac{p_{\alpha\,i}^2}{2m_\alpha}
+\sum_{\alpha\,i\neq\beta\,j} \left[ 
\frac{Q_{\alpha\,i}^a\,Q_{\beta\,j}^a}{|x_{\alpha\,i}-x_{\beta\,j}|}
+V_{core}\right] \, ,
\label{HCB}
\ee
where the sum is over the species $\alpha=q,\overline{q},g$ and their
respective numbers $N_\alpha$, each carrying a thermal energy
$m_\alpha\approx \pi T$. We have added a short-range repulsive core, 
which is needed to regulate the short-distance integrals and dynamical 
simulations~\footnote{Minimally this term is caused by quantum
  localization energy.}.
The phase space coordinates are position ($x_\alpha$), momentum ($p_\alpha$) 
and color ($Q_\alpha$). Only the Coulomb-like interaction was retained in 
(\ref{HCB}) in the non-relativistic limit. The non-perturbative effects due
to chromomagnetism will be discussed elsewhere.

A central issue in {\it classical colored} plasmas is screening. In the weak 
coupling limit the linear Debye-H$\ddot{\rm{u}}$ckel theory is well 
established  and we will
describe it below. At intermediate couplings (say $\Gamma=1-10$ in the
notations of I) not much is known and we will give some suggestions in
this direction below. At large couplings (say $\Gamma=100-200$ in the
notations of I) a more adequate description can be developed starting
instead from the crystal limit.

\section{Classical QGP Partition Function}

In this section we discuss the thermodynamics of cQGP.
We formulate a generalized partition function for the 
cQGP to all orders in the classical coupling $\Gamma$
(ratio of the potential to kinetic energy) and density. We
discuss screening as a mean-field phenomenon. We first discuss 
its weak coupling limit in the form of a linear
Debye-H$\ddot{\rm{u}}$ckel limit, and corrections through a low density 
virial expansion.

The partition function of the cQGP is

\begin{eqnarray}
&& Z_N=\int\,\prod_\alpha\frac 1{N_\alpha !}\,
\prod_{\alpha\,i}^{N_\alpha}\frac{dQ_{\alpha\,i}\,dx_{\alpha\,i}}
{\lambda_\alpha^3}\,\nonumber\\
&&\times{\rm exp}(-\frac 12\int\,dx\,dx'\,
\rho^a(x)\,v(x-x')\rho^a(x') \nonumber\\
&&\qquad\qquad+\rho(x)\,w(x-x')\,\rho(x'))
\label{DH}
\end{eqnarray}
with the color charge density

\be
\rho^a(x)=\sum_{\alpha\,i}Q_{\alpha\,i}^a\,\delta(x-x_{\alpha\,i})
\label{DH1}
\ee
and the density

\be
\rho(x)=\sum_{\alpha\,i}\,\delta(x-x_{\alpha\,i})
\label{DH2}
\ee
The unscreened and dimensionless Coulomb potential is
$v(x)=l_B/r$ with $l_B=g^2/(4\pi\,T)$
the length at which two charges interact with
energy $T$~\footnote{We have substituted $g^2$ by $g^2/4\pi$ to
make contact with standard SI conventions in strongly coupled plasmas}.
To prevent a classical collapse of the cQGP,
we introduce a phenomenological {\rm repulsion} with a core
$a$ that is $w(x)=\infty$ for $x<a$ and zero otherwise.
In a quantum theory $a$ is set by the Heisenberg uncertainty
principle. The $\lambda_\alpha$'s in (\ref{DH}) are just the
thermal wavelengths for each species $\alpha=q,\overline{q},g$.

At weak coupling or in the gas phase, the partition function
(\ref{DH}) is expected to scale as

\be 
Z_N^{\it gas}\approx  (\lambda\,T)^{(3/2+\alpha(N_c))\,N} 
\label{Z1}
\ee 
The $3/2$ is standard from the Maxwellian distribution over
momenta, while $\alpha(N_c=2)=1$ and $\alpha(N_c=3)=3$ are 
extra powers from the classical color degrees of freedom. 
Indeed for $N_c=3$ there are 3 angles and 3 conjugate
angles (generalized momenta) in addition to the 2 fixed
Casimirs that make the 8-dimensional classical color vector.

At very strong coupling or in the crystal phase, the partition
function (\ref{DH}) is expected to scale as

\be 
Z_N^{\it crystal}\approx (\lambda\,T)^{(3+\alpha(N_c))\,N} 
\label{Z2}
\ee
since the crystal localizes the particles in space as well.
The change in behavior between (\ref{Z1}) and (\ref{Z2}) 
is reflected in the change in the specific heat of each phase.
The partition function and thus the specific heat of the 
liquid is intermediate between (\ref{Z1}) and (\ref{Z2}).

\subsection{Mean-field screening}

The grand-partition function follows from (\ref{DH}) through

\be
Z_\lambda=\sum_{N}\lambda^{N_q+N_{\overline{q}}+N_g}\,Z_N
\label{GDH}
\ee
where the classical fugacities for the quarks, antiquarks and gluons
are the same as the three species are all in the adjoint representation
with comparable thermal energies $m_\alpha$.

Since (\ref{DH}) is Gaussian 
in the densities it can be linearized using standard Hubbard-Stratonovich 
transformations. The result after exponentiation is 

\be
Z_\lambda=\int\,\frac{d\vec\phi\,d\psi}{Z_vZ_w}e^{-{\bf S[\vec\phi , \psi]}}
\label{ZZ}
\ee
with the induced action

\be
{\bf S}[\vec\phi ,\psi]=
&&\frac 12\int\,dx\,dx' \phi^a(x)\,v^{-1}(x-x')\phi^a(x') \nonumber\\
&&+\frac 12 \int\, dx\,dx' \psi(x)\,w^{-1}(x-x')\,\psi(x')\nonumber\\
&&+\lambda\sum_\alpha\,dQ_\alpha\,dx_\alpha\nonumber\\
&&\times\,e^{\frac 12 (w(0)+v(0)+Q_\alpha^2
-iQ_\alpha^a\phi^a(x_\alpha)-i\psi(x_\alpha))}
\label{SA}
\ee
The contributions $v(0)$ and $w(0)$ are the divergent self-energies. 
The normalizations $Z_{v,w}$ are

\be
Z_v=&&\left({\rm det}v\right)^{(N_c^2-1)/2}\nonumber\\
Z_w=&&\left({\rm det}w\right)^{1/2}
\ee
This analysis borrows on the functional framework developed
for various ionic systems as discussed in~\cite{SOLUTES}
(and references therein).

The induced action (\ref{SA}) captures the essentials of the cQGP at
all couplings. It is instructive to see that the saddle point 
approximation $\delta{\bf S}/\delta\vec{\phi}=0$ with $\psi=0$
yields

\be
\nabla^2\vec\phi & =&4\pi\,l_B\,\sum_\alpha\int\,dQ_\alpha\,
i\,\vec{Q}_\alpha\,(\lambda_\alpha\,e^{Q_\alpha^2\,v_0/2})\nonumber\\
&\times&\left(e^{-i\vec{Q}_\alpha\cdot\vec\phi}-1\right)
\label{MF1}
\ee
If we define $\vec\phi=(-ig/k_BT)\,\vec\varphi$, then
(\ref{MF1}) reduces to

\be
\nabla^2\vec\varphi & =& -\sum_\alpha\, g\,c_\alpha \nonumber \\
&& \int \,dQ_\alpha\,\vec{Q}_\alpha\,
\left(e^{-g\vec{Q}_\alpha\cdot\vec\varphi/k_B\,T}-1\right)
\label{MF1X}
\ee
which is just the {\it non-linear} 
Debye-H$\ddot{\rm{u}}$ckel equation for the cQGP with 
the renormalized concentration

\be
c_\alpha=\lambda_\alpha\,e^{Q_\alpha^2\,v_0/2}\,\,.
\ee
For one species, a trivial mean-field solution is $\vec\varphi=
\vec{Q}\varphi$ with $\nabla^2\varphi=0$.

The mean-field equation (\ref{MF1X}) is valid for both weakly
and strongly coupled plasma. At weak coupling it reduces to the
linear Debye-H$\ddot{\rm{u}}$ckel theory which we will explore below. 
At intermediate and strong coupling  it corresponds to the non-linear 
Debye-H$\ddot{\rm{u}}$ckel description
of which the crystalline structure is an asymptotic limit. This will
also be discussed below.

\subsection{Linear Debye-H$\ddot{\rm{u}}$ckel Screening}

The linearized Debye-H$\ddot{\rm{u}}$ckel limit follows from (\ref{MF1X}) for 
weak coupling or high temperature, whereby the exponent is 
expanded to first order leading to

\be
(\nabla^2+K^2)\vec\varphi \approx 0
\ee
This suggests to organize (\ref{SA}) around the 
Debye-H$\ddot{\rm{u}}$ckel limit. 
Specifically, we introduce the screened Coulomb potential

\be
v_{DH} (x)= \frac{l_B}r\, e^{-Kr}
\label{DHH}
\ee
with $K^2=4\pi\,l_B\,\lambda\sum_\alpha Q_\alpha^2/(N_c^2-1)$. 
To leading order $\lambda$ is the concentration of tertiaries

\be
\lambda\approx c= \frac{N_q+N_{\overline{q}}+N_g}{3V}
\ee
as we will show below. So $K^2$ is just the sum of the second Casimir's
weighted with the species concentration. Using the inverse relation

\be
v_{DH}^{-1}=v^{-1}+\frac{K^2}{4\pi\,l_B}
\label{INV}
\ee
We may reorganize (\ref{ZZ}) around the linearized
Debye-H$\ddot{\rm{u}}$ckel limit. Specifically,

\be
Z_\lambda=
e^{3\lambda\,V+\frac 12\lambda Vv(0)\sum_\alpha Q_\alpha^2}
\,\frac{Z_{DH}}{Z_v}\,\langle e^{\lambda\sum_\alpha{\bf K}_\alpha}
\rangle
\label{ZZZ}
\ee
with

\be
Z_{DH}=\left({\rm det}v_{DH}\right)^{(N_c^2-1)/2}
\ee
and

\begin{eqnarray}
{\bf K}_\alpha=&&\int\,dQ_\alpha\,dx_\alpha\,\nonumber\\
&&\times(e^{\frac 12 (w(0)+v(0)+Q_\alpha^2
-iQ_\alpha^a\phi^a(x_\alpha)-i\psi(x_\alpha))}\nonumber\\
&&-1-\frac 12 v(0)\,Q_\alpha^2+\frac 12\frac{Q_\alpha^2}{(N_c^2-1)}\,
\phi^a(x_\alpha)\phi^a(x_\alpha))
\end{eqnarray}
The averaging in (\ref{ZZZ}) is carried using the linearized
Debye-H$\ddot{\rm{u}}$ckel measure

\be
\frac 1{Z_{DH}Z_w}e^{-\frac 12 \phi^a\,v_{DH}^{-1}\,\phi^a
-\frac 12\psi\,w^{-1}\,\psi}\,\,\,.
\ee
We note that (\ref{ZZZ}) while organized around the linearized
Debye-H$\ddot{\rm{u}}$ckel limit, it still includes the non-linear corrections
in the coupling to all orders.

In the linearized Debye-H$\ddot{\rm{u}}$ckel limit, 
the dimensionless pair interaction is 

\be
g(1,2)\approx e^{-{\bf Q}^a_1{\bf Q}_2^a\,v_{DH}(R_1-R_2)}
\ee
and fulfills the superposition principle in liquids

\be
g(1,2,3)=g(1,2)\,g(2,3)\,g(1,3)
\label{sup}
\ee
Also, (\ref{INV}) is just the Ornstein-Zernicke equation in liquids

\be
v(x)=v_{DH} (x)+\frac{K^2}{4\pi l_B}\int dx'\,v_{DH} (x') \,v(x-x')
\label{OZ}
\ee
(\ref{ZZZ}) allows us to go beyond this equation 
systematically. The correlations introduced by ${\bf K}\neq 0$ 
upset the superposition principle (\ref{sup}) and
the Ornstein-Zernicke equation (\ref{OZ}).

\subsection{Virial Corrections to Linear Debye-H$\ddot{\rm{u}}$ckel Screening}

At low density or small $\lambda$ we can systematically
correct the linearized Debye-H$\ddot{\rm{u}}$ckel limit. The purpose
of it is to see how stable is the linearized theory for
higher densities. We note that
an expansion in $\lambda$ of $e^{\lambda\,{\bf K}}$ in
(\ref{ZZZ}) is {\it not} an expansion in the coupling
(although the diluteness is a property of the weakly
coupled phase). Thus, we define the rescaled 
and dimensionless variables using the short distance cutoff
(core size) $\tl =a^3\lambda$, $\tc =a^3\,c$ and $\tv =V/a^3$.
The rescaled free energy is just

\be
{\bf\tilde F}=3\,\tc\,{\rm ln}\,\tl -\frac 1{\tv}
{\rm ln}\,{Z_\lambda}
\label{FE}
\ee
and the rescaled concentration per species is

\be
\tc =\frac 1{3\tv} \frac{\partial {\rm ln}{Z_\lambda}}{\partial{\rm ln}\tl}
\label{RC}
\ee
The first contribution in (\ref{FE}) is that of free particles
while the second contribution is the excess/deficit free energy.
The excess/deficit is caused by the interaction. 

The excess/deficit free energy can be virial expanded for low
concentrations whatever the coupling,

\be
\frac{a^3{\rm ln}Z_\lambda}{V}=
3\tl - B_{3/2}\tl^{3/2}-B_2\tl^2-B_{5/2}\tl^{5/2}+...
\label{V1}
\ee
in terms of which the concentration reads

\be
\tc=\tl-\frac{1}{2} \,B_{3/2}\tl^{3/2}-\frac 23\,B_2\tl^2-
\frac 56\,B_{5/2}\tl^{5/2}+...
\label{V2}
\ee
The latter can be inverted to give

\be
\tl =&&\tc+\frac{1}{2} \,B_{3/2}\tc^{3/2}+\frac 23(B_2+\frac 9{16}B_{3/2}^2)
\tc^2\nonumber\\
&&+\frac 23(\frac 54B_{5/2}+\frac 74 B_2B_{3/2}+\frac {69}{153}B_{3/2}^3)
\tc^{5/2} +...
\label{V3}
\ee
In terms of the concentrations, the free energy is then

\be
{\bf\tilde F}=3\tc\,{\rm ln}\,\tc-3\tc+
D_{3/2}\tc^{3/2}+D_2\tc^2+D_{5/2}\tc^{5/2}+...
\label{V4}
\ee
with 

\be
&&D_{3/2}=B_{3/2}\nonumber\\
&&D_2=B_2+\frac 38 B_{3/2}^2\nonumber\\
&&D_{5/2}=B_{5/2}+B_2B_{3/2}+\frac 7{32}B_{3/2}^3
\label{V5}
\ee

The virial expansion for arbitrary $N_c$ is involved.
For $N_c=2$ all color integrations
can be done explicitly. The result is

\be
D_{3/2}=-\sqrt{4\pi}\left(\sum_\alpha\epsilon_\alpha\right)^{3/2}
\ee
and

\begin{eqnarray}
D_2=&&\frac{3\pi}{2}\left(\sum_\alpha\epsilon_\alpha\right)
\left(-3\sum_\beta\epsilon_\beta^2+(\sum_\beta\epsilon_\beta)^2\right)
\nonumber\\
&&-\frac{\pi}{24\epsilon}\sum_{\alpha,\beta}
(54\epsilon^4\int_0^{3\epsilon}\,\frac{{\rm sh}t}t+\nonumber\\
&&\qquad e^{-3\epsilon}(2-2\epsilon-3\epsilon^2-9\epsilon^3)\nonumber\\
&&\qquad +e^{3\epsilon}(-2-2\epsilon+3\epsilon^2-9\epsilon^3))
\end{eqnarray}
and

\be
B_{5/2}= \frac{54\pi^{3/2}}5
\left(\sum_\alpha\,\epsilon_\alpha\right)
\,\left(\sum_\alpha\epsilon_\alpha^2\right)^2\,
{\rm ln}\,{\Lambda\,a}
\label{div}
\ee
with the classical binding energy

\be
\epsilon_\alpha=C_{2\alpha}\frac{l_B}a
\ee
and $\epsilon=\sqrt{\epsilon_\alpha\,\epsilon_\beta}$.
In (\ref{div}) only the logarithmic contribution was kept with 
the dimensionless cutoff $\Lambda\,a$ reflecting on the infrared
sensitivity of the virial coefficient.

$D_{3/2}$ is the linear Debye-H$\ddot{\rm{u}}$ckel contribution to the
excess free energy while $D_2$ is the correction due to  2-body effect
in the screened phase. The contributions $e^{\pm 3\epsilon}$
are the Boltzmann contributions from would-be molecules made of 2
charges a distant $a$ apart. The virial expansion
is exponentially sensitive to molecule formation, pointing to the
quantum character of the sQGP even at low concentrations, i.e.

\be
\tc_*\approx \left|\frac{D_{3/2}}{D_2}\right|^2
\approx \left(\frac{48}{9\sqrt{\pi}}\right)^2
\left(\frac{e^{-6\epsilon}}{3\epsilon}\right)
\ee
For a 2-color sQGP, the Casimirs 
$C_{2q}=C_{2\overline{q}}=1/2$ and $C_{2g}=3$
are of order 1. Setting $C_2\,\alpha_s\approx 1$
yields $\epsilon\approx 1/aT$. For a core $a\approx 1/2T$,
the virial expansion around the linearized Debye-H${\ddot{\rm u}}$ckel
limit breaks down for concentrations as low as
$c_*\approx 10^{-4}T^3$.

\section{Nonlinear Poisson-Boltzmann Screening}

To go beyond the linear Debye-H$\ddot{\rm{u}}$ckel theory we need to solve
the full mean-field equation (\ref{MF1X}),
known as a  Poisson-Boltzmann equation. In this section, we 
do so using the
simplifying assumption that the non-linear solution remains spherically
symmetric. This allows us to probe the stability of the linearized theory
at strong coupling or higher densities. However, the reader should be 
prepared to see that at sufficiently strong coupling it would not be 
adequate either, with correlated charges forming non-spherical crystalline 
order.

The non-linear radially symmetric Debye-H$\ddot{u}ckel$
equation for the {\it Abelian} potential is

\be 
\phi''(r)+2\phi'(r)/r=4\pi\delta n (\phi)
\label{eqn_mf1}
\ee
where the density variation in the r.h.s. can be written as a
Boltzmann exponent\footnote{The fact that we use the same $\Gamma$ in
both the Coulomb and core terms means that an increase in the 
coupling actually means a reduction of the
temperature.} of the potential
\be 
\delta n (\phi)=n\left( e^{\Gamma(\phi-V_{core})}-1\right) \, ,
\label{eqn_mf2}
\ee
where the first term in the exponent is the Coulomb potential, 
defined to be positive. The second term, the core, is now needed
to ensure that the density variation associated with one particle
is normalizable at small $r$. Below we set the normalization as

\be \label{eqn_norm} \int d^3r \,\delta n (\phi)=1 \ee

\begin{figure}[ht]
\begin{center}
\epsfig{figure=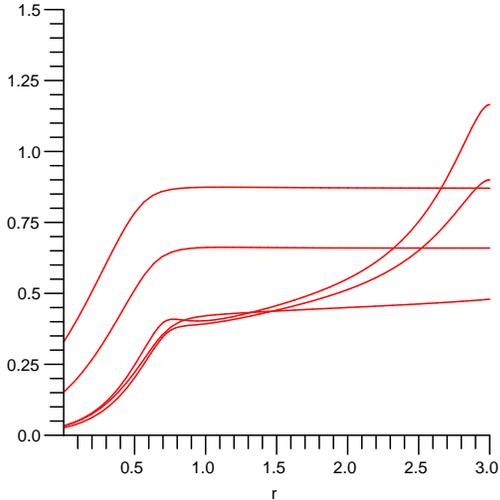,width=8cm,width=8cm}
\epsfig{figure=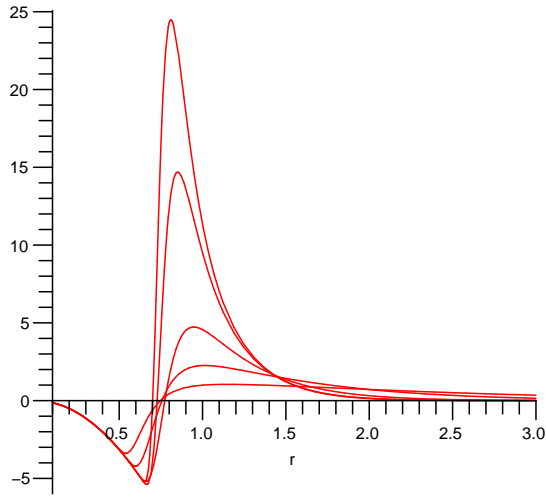,width=8cm,width=8cm}
\bigskip
\caption{(Color online) (a) The function $f(r)$ 
for $\Gamma=0.1,0.3,1,2,2.2$ (top to bottom at the l.h.s.).
(b) The integrand $4\pi\, r^2\, \delta n(r)$ for the 
same values of the coupling. Larger couplings produce higher peaks.}
\label{fig_strongscreening}
\end{center}
\end{figure}

We set $n=1$ below and seek a radial solution of the form

\be  
\phi(r)= f(r)\,\frac{e^{-\kappa r}}r
\label{radial}
\ee
with $\kappa=\sqrt{4\pi\Gamma}$ the Debye-H$\ddot{\rm{u}}$ckel inverse
radius. One may think of $f(r)$ as of an effective coupling.
The value $f(r)=1$ corresponds to
the normalized solution of the linearized theory. If one includes
the repulsive core at small distances, in weak coupling 
f(r) remains an r-independent constant, although different
from 1. As we will see, at stronger coupling the deviations
of f(r) from a constant value will display deviations from
the linear Debye-H$\ddot{\rm{u}}$ckel theory.

Inserting (\ref{radial}) into (\ref{eqn_mf1}) allows for numerical
solutions for weak and strong coupling. At large distances, the
screened potential is weak and the linear Debye-H$\ddot{\rm{u}}$ckel 
theory is
valid with $f(r)$ asymptotic a constant. The latter is fixed by
the normalization (\ref{eqn_norm}). The numerical results for $f(r)$ 
for different $\Gamma$ are shown in Fig.~\ref{fig_strongscreening}(a). 
As expected, for $\Gamma<1$ the function $f(r)$ asymptotes a constant,
except at small distances where the linearized theory never works.
For $\Gamma>2$ one enters a domain where $f(r)$ 
is changing dramatically.  We found that for $\Gamma>2$ the modification is fast growing,
indicating a beginning of a phenomenon known as 
``over-screening''\footnote{
After the work was basically completed the authors learned that it is
 widely used  in important
 chemical and biological applications, see review
in ref.\cite{Shklovskii}.}. At large $\Gamma$ our solution is  
large and
oscillating, which indicate not only large deviations from a Debye
theory but actually 
a complete breaking of the mean field 
approach itself. Such erratic solutions
are  a precursor for highly correlated state, beyond spherical
mean field,
and eventually a complete crystallization at very large $\Gamma$ .

The generalization of these results to a {\it non-Abelian} colored plasma is
straightforward. In the $SU(2)$ case, the
original point charge can be thought of as having a particular color
direction, while those in a screening cloud have any charges.
As a result, there is an angular variable 
$z={\rm cos}(\theta)=\hat{Q}\cdot\hat{Q}_0$
in the Boltzmann factor as in (\ref{MF1X}). The
potential $\Phi_{SU2}$ averaged over the color orientations is
\be  
\Phi_{SU2}=\frac 12
\int_{-1}^{1} dz\, z\, e^{zA} 
= \frac{{\rm ch}A}{2A}-\frac{{\rm sh}A}{2A^2}
\label{SU2}
\ee
where $A=\Gamma\Phi$ is the Coulomb part. The Casimir is
absorbed in $\Gamma$. These factors are reminiscent of 
the $e^{\pm 3\epsilon}$ in the virial expansion and have
the same origin. (\ref{SU2}) is plotted
in Fig~\ref{fig_colorscreening} along with its asymptote. 
The linear {\it non-Abelian} mean-field solution follows from 
the {\it Abelian} one by multiplying the squared Debye mass by
$<z^2>=1/3$. Other gauge groups can be
treated in the the same way. Indeed, in the $SU(3)$ case the averaging
over angles include 6 variables (the 8 generators minus the 2 Casimirs)
in the pertinent invariant measure.

\begin{figure}[t]
\begin{center}
\epsfig{figure=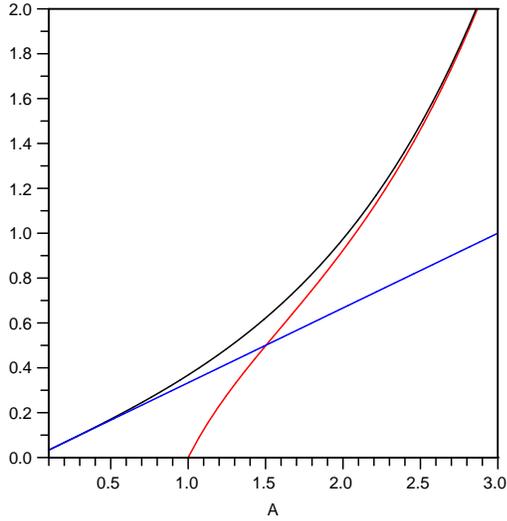,width=8cm}
\bigskip
\caption{(Color online) $\Phi_{SU2}(r) $ (upper curve) with its small
$A/3$ and large $e^A\,(A-1)/2A^2$ expansion.
}
\label{fig_colorscreening}
\end{center}
\end{figure}

\section{Nonlinear Screening through a Coulomb Crystal}

As we pointed out before, at very large  $\Gamma$
or very small $T$  the system freezes into a crystal. 
To understand the difference between screening in the 
Debye-H$\ddot{\rm{u}}$ckel
limit and the screening in the crystalline phase, we discuss the
induced potentials and the behavior of their pertinent partition
functions.

In the Debye limit the induced potential at small
and intermediate coupling is of the form

\be 
V_D(r)=-{Q\over r}\left(1-e^{-{r/R_D}}\right)
\approx -{Q\over R_D}\left(1-{r/2R_D} + ...\right)
\ee 
The induced potential is finite at small $r$ and rises linearly
with $r$. The pertinent classical partition function in the
Einstein approximation (ignoring coupling between oscillations
of different charges) is

\be 
Z\approx \int d^3 r \, e^{-{V_D/T}}
\approx T^3 \, e^{Q/T\,R_D} \ee

In a crystal the induced potential by all charges except one
is quite different. If one starts with a pure Coulomb field
in a cubic crystal, then

\be V(\vec r)= 
\sum_{n_i=-n_{max}}^{n_{max}}{ (-1)^{\sum n_i}\over 
|\vec r - \vec n|}-{1\over |\vec r| }
\ee
where we introduced a cutoff of the sum $n_{max}$ and also subtracted
the $n_x=n_y=n_z=0$ term. The alternating signs cause the sum to converge.
To smoothen-out the oscillations between even-odd finite $n_{max}$ we use
instead

\be 
\bar V(x,y,z)= {V(n_{max})+V(n_{max}+1)\over 2} 
\ee
which is within few percent of the limit already at $nmax=3$.
The convergence is related to the fact that new distant charges
added between $n_{max}$ and $n_{max}+1$ do not
have a nonzero dipole or quadrupole contribution.
In fact they start contributing to the l=4 octupole. As a result,
the field is extremely flat at small $r$ (near the origin)
with only quartic $r^4$ corrections to the constant $V(0)$,
see Fig.\ref{fig_Coulomb}. This is just Madelung constant
which is the energy of the ideal zero $T$ crystal (in our
case it is 1.75). The figure shows also high multiples
in the angular distribution, with alternating 
positive and negative (unstable) directions.

\begin{figure}[h]
\begin{center}
\epsfig{figure=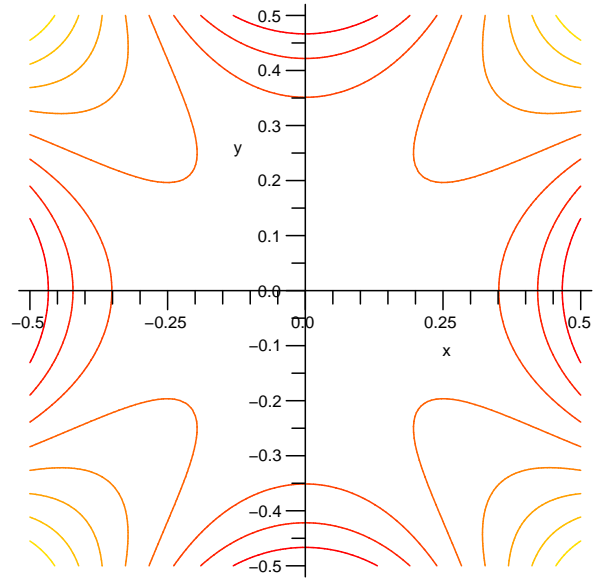,height=8cm,width=8cm,angle=0}
\bigskip
\caption{(Color online)
Contour plot of the Coulomb potential in the z=0 plane versus
  x and y (in units of the nearest neighbor distance) as
induced by all charges except one in a cubic crystal.}
\label{fig_Coulomb}
\end{center}
\end{figure}

We note that the potential used in our molecular simulations 
in \cite{GSZ_I} has in addition a {\it repulsive core} needed
to prevent collapse at short distances. The result is a 
quadratic instead of quartic potential at the origin as is
seen in Fig.~\ref{fig_core}, with $V(r)\approx 9r^2$ near
the origin. The pertinent partition function in the 
Einstein approximation is then

\be 
Z\approx \int d^3 r \, e^{-{V/ T}}
= \left ({\pi T \over 9}\right )^{3/2} \, ,
\ee
which leads to the usual classical potential energy of oscillations
$3T/2$. However, as we have shown above, in order to understand the MD
data one has to include the oscillations of the color vectors as well.

\begin{figure}[h]
\begin{center}
\epsfig{figure=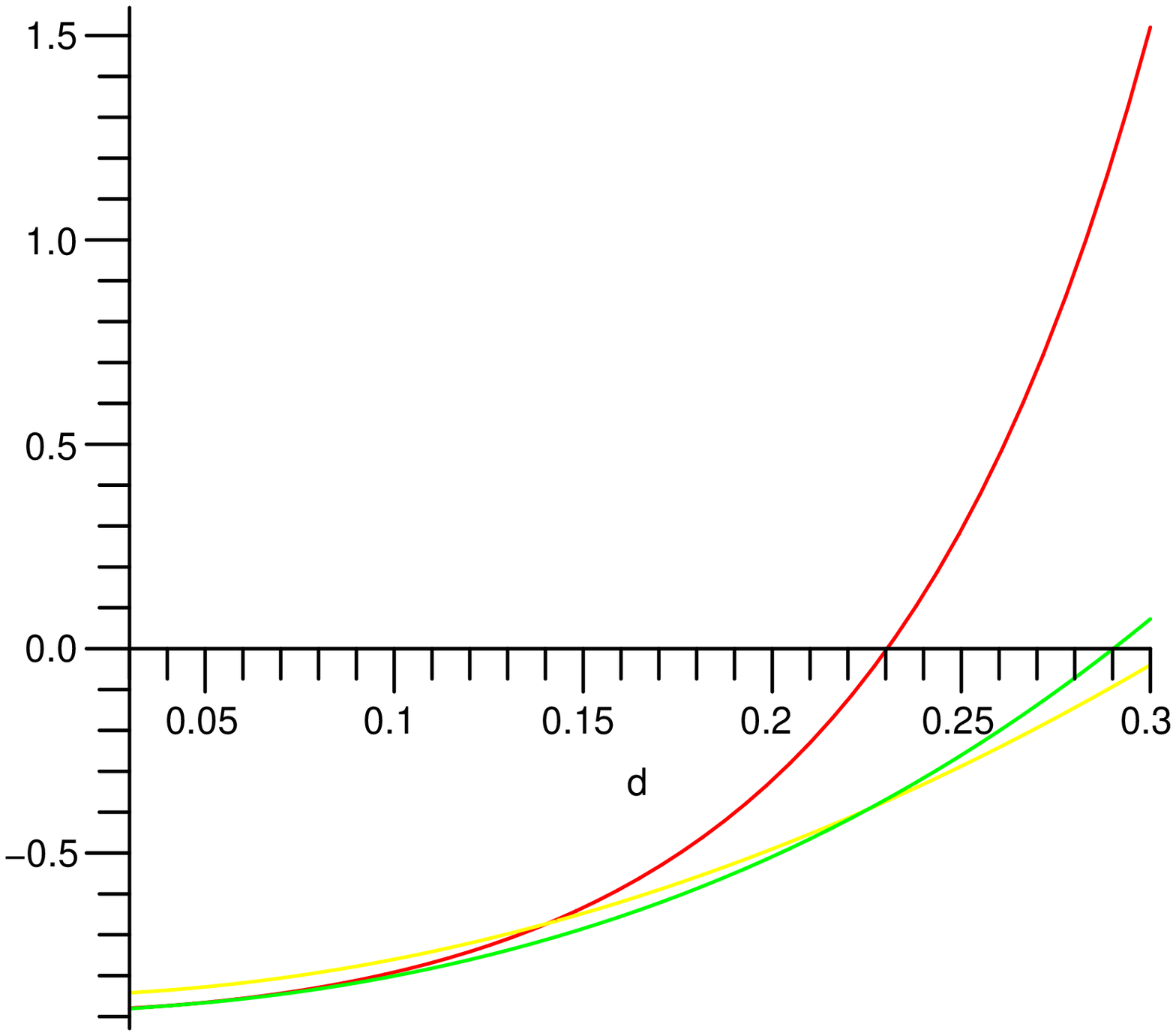,height=6cm,width=6cm,angle=0}
\epsfig{figure=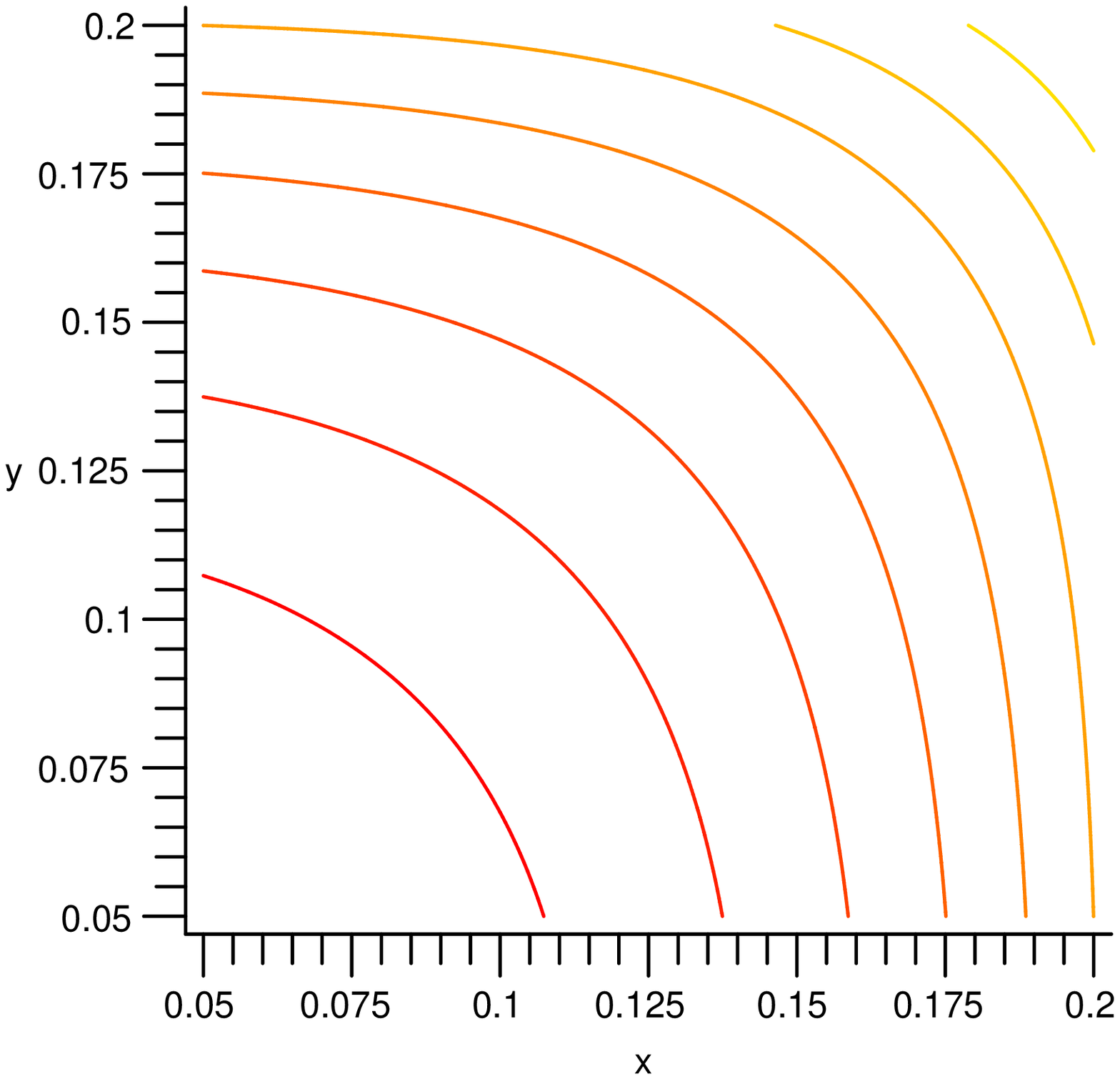,height=6cm,width=6cm,angle=0}
\bigskip
\caption{(Color online) 
Induced potential with Coulomb and repulsive core
in the z=0 plane: (a) along the crystal axes (green) and the diagonal
(red), along with the fit (yellow). (b)  Planar contour plot for a cubic 
crystal.}
\label{fig_core}
\end{center}
\end{figure}

\section{From Debye-H$\ddot{\rm{u}}$ckel Screening to Debye Crystal}

As the density is increased or equivalently the coupling
is increased or the temperature is decreased, the potential effects
dominate over the kinetic effects in the cQGP. $\Gamma$ which
is the ratio of the potential to kinetic energy is then large.
For simplicity and to compare with the molecular dynamics results in I, we
specialize in this section to one species, say gluons only. Let $\Gamma$
be the standard cQGP plasma parameter defined as~\cite{GSZ_I}

\be
\Gamma=\frac{\rm potential}{\rm kinetic}=\frac{C_2\alpha_s/a_{WS}}{T} \, ,
\label{gamma}
\ee
with the Wigner-Seitz radius satisfying $c\,(4\pi\,a^3_{WS}/3)=1$.
In terms of (\ref{gamma}) the excess free energy $\Fex$ is

\be
{\bf F}(\Gamma)={\bf F}(0) +\Fex(\Gamma)
\ee
and measures the Coulomb interacting part of the free energy.
In the cQGP the excess free energy $\Fex$ per particle 
is the Debye-H$\ddot{\rm{u}}$ckel contribution

\be
\frac{\Fdh}{N\,T}=&&(N_c^2-1)\,\frac 12\int\frac{d^3q}{(2\pi)^3}
\left({\rm ln}\left(1+\frac{K^2}{q^2}\right)-\frac{K^2}{q^2}\right)
\nonumber\\
=&&-(N_c^2-1)\frac{\Gamma^{3/2}}{\sqrt{3}} \, ,
\label{DH*}
\ee
which is finite, attractive  and nonlinear in the concentration $c$. 

The higher order corrections in the concentration and/or coupling were
discussed above and will not be repeated. Instead, here we note that
(\ref{DH*}) is finite and involves an integration over small as
well as large momenta. As the density is increased in the cQGP there is
increasing local ordering due to the strong colored Coulomb attractions
causing the dilute and screened gas to undergo changes to a liquid and
eventually to a crystal.

Following Debye description of solids, we will
assume that the allowed momenta in (\ref{DH*}) are those commensurate
with the total number of degrees of freedom,

\be
2\int_0^{q_D}\frac{V\,d^3q}{(2\pi)^3}=3\,N\,d
\label{QD}
\ee
which fixes the Debye cutoff $q_D=\pi\,(9d\,c)^{1/3}$. 
For every particle moving in 3d space there are attached $d=N_c(N_c-1)/2$
independent but classical color charges. The cutoff Debye-H$\ddot{\rm{u}}$ckel
contribution is

\begin{eqnarray}
&&\frac{\Fdh}{N\,T}=(N_c^2-1)\,\frac{q_D^3}{8\pi^2}\nonumber\\
&&\times\,\int_0^1\,x^2\,dx\,
\left({\rm ln}\left(1+\frac{A\,\Gamma}{x^2}\right)-\frac{A\,\Gamma}{x^2}\right)
\label{CRYS1}
\end{eqnarray}
The integrals are readily undone and the answer is

\begin{eqnarray}
&&\frac{\Fdh}{N\,T}=-(N_c^2-1)\,\frac{3d}{2}\nonumber\\
&&\times\,\left((A\Gamma)^{3/2}\,{\rm atan}(1/\sqrt{A\Gamma})-
\frac 12({\rm ln}(1+A\Gamma)-A\Gamma)\right)
\label{CRYS2}
\end{eqnarray}
with $A=(2/3)(2/(\pi\,d)^2)^{1/3}$. 
A similar construction was suggested by Brilliantov 
for the Abelian one component plasma~\cite{BRILLIANTOV}.
The excess energy following from (\ref{CRYS2}) is
${{\bf U}}_{\rm ex}=\partial\Fdh/\partial{\rm ln}\Gamma$
which is

\be
\frac{{\bf U}_{\rm ex}}{N\,T}=-(N_c^2-1)\,
\frac {9d}4 \left((A\Gamma)^{3/2}\,{\rm atan}(1/\sqrt{A\Gamma})\right)
\label{UEXCESS}
\ee
For small $\Gamma$ we recover the (linear)
Debye-H$\ddot{\rm{u}}$ckel limit, while
for large $\Gamma$  we obtain 

\be
\frac{{\bf U}_{\rm ex}}{N\,T} & = &-(N_c^2-1)\,\frac{9d}{4}
\left(A\, \Gamma - {1\over 3} \right) + {\cal O}(\Gamma^{-1})
\nonumber \\
& = & - c_{MD}\,\Gamma + {9\over 4}
\label{MADEL}
\ee
where the Madelung constant is $c_{MD}$.  
For $N_c=2$: $d=1$, $A\approx 0.392$ and 
$c_{MD}=2.64$

\vspace{0.3in}

\begin{figure}[ht]
\begin{center}
\epsfig{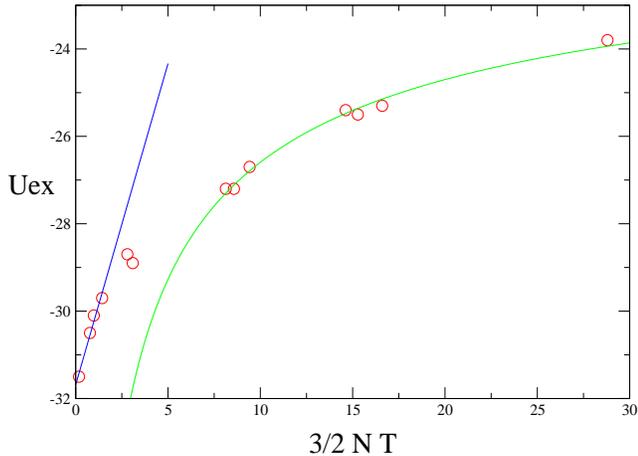}
\bigskip
\caption{(Color online)
Potential energy as a function of kinetic energy, $(3/2) N T$ for N=64
 particles is N=64. The red points are the MD simulations,
their statistical accuracy (not shown) is estimated to be 7 percents
for all points.
The blue line is a fit to eq.~(\ref{linear_fit}).
the green curve is a fit to eq.~(\ref{DH_fit}).}
\label{uex}
\end{center}
\end{figure} 

This excess potential energy was calculated
numerically in I for the one-species cQGP with $N_c=2$ 
using molecular dynamics. Figure~\ref{uex} shows the plot of the excess 
potential energy as a function of the total kinetic energy, $(3/2) N T$.
The region of small temperature corresponds to large $\Gamma$ while the
region with large temperature to small $\Gamma$. For points corresponding
to $\Gamma \approx 16 - 131$ the excess energy can be approximated 
by~\footnote{See Appendix for the discussion of units.} 

\be
{{\bf U}_{ex} \over N} \approx -0.5 + 2.25 \, T \, .
\label{linear_fit}
\ee

The numerical crystal analysis of section V yields
the following excess energy per particle 

\be
{{\bf U}_{ex} \over N} \approx -0.44 + 2.5 \, T \, 
\ee
with $3T/2$ following from the oscillation energy 
in the harmonic potential, plus $1\, T$ from the oscillations
of the classical $SU(2)$ color degrees of freedom~\footnote{As explained in I,
the unit color vector on a sphere is equivalent
to an oscillator with one coordinate and one momentum.}.
One can see that at least the specific heat of a low-T colored crystal
is well reproduced.

Using the definition of $\Gamma$ in (\ref{gamma}) in (\ref{MADEL})
the excess energy per particle following from the 
Debye interpolation as a function of temperature is

\be
{{\bf U}_{ex} \over N} \approx -0.65 + 2.25 \, T .
\ee
which is in in good agreement with the molecular dynamics result
at small temperature or large  $\Gamma$. In the dimensionless
units of the Appendix, the conversion from $T$ to $\Gamma$
is given by $\Gamma=0.25/T$.

In the intermediate coupling regime with $\Gamma \approx 0.8 - 3$,
we enter the liquid regime of the cQGP (the right part of the figure
as small $\Gamma$ means high $T$). Numerically,

\be
{{\bf U}_{ex} \over N} \approx -0.3 - {0.03 \over \sqrt{T}}
\label{DH_fit}
\ee
The Debye-Huckel limit predicts ${\bf U}_{ex}/N=-0.3/\sqrt{T}$ at large $T$ 
or weak coupling but not the large {\it negative} constant.
Its appearance can be traced back to the appearance of a
sharp peak in the inter-particle distance at about $1/n^{1/3}$
as shown in Fig.1b in the liquid phase. In the gas phase, the
the peak is spread in the form of a large {\it Debye cloud}.
The negative constant contribution to the energy in (\ref{DH_fit})
corresponds to the potential following from this peak.

\section{Summary}

We have analyzed the screening and
thermodynamics of a strongly coupled classical colored QGP.
The basic question we asked in this paper was when and whether one can
relate the textbook theories, such as Debye-H$\ddot{\rm{u}}$ckel screening or
virial expansion, to our MD data.

We have identified the onset of nonlinear screening
using the mean field approach, as well as onset of the correlations
using a low density expansion around the mean field. We
have carried explicitly the expression for the pressure of the cQGP
to order $c^{5/2}$. Not unexpectedly, we have found
that the textbook methods are not really applicable
 in the liquid regime we studied.
In particular, the spherically symmetric mean field treatment 
is very unstable already at medium coupling, and the virial expansion
fails at very small 
densities. Both are signals for the onset of a cluster formation
(liquid) or long range ordering (crystal).

We hope to address more issues in forthcoming publications.
In particular, issues related to the 
collective excitations of the system (sound and
plasma-color waves), and
propagation of external bodies in QGP (jet energy
loss) etc.  

\section*{Appendix}

In this appendix we briefly discuss the units that were used in I:
length, time and mass. The unit of length was chosen as
the separation between two particles, $r=\tilde\lambda$,  at which the 
potential (the sum of Coulomb and core) is minimum. 
The density or concentration of particles $n=c$, 
is dimensionless in these units.
All simulations in I were carried with $n=c=0.3676$.
The unit of time $\tau_0$ was set by the plasma frequency

\be 
\tau_0=\omega_p^{-1}= ({m\over 4\pi n e^2})^{1/2} \, ,
\ee
with $e^2=C_2\alpha_s$ in the non-Abelian case.
The unit of mass was set by the particles energy $m$.

All dimensional quantities can be expressed using 
these basic units. For instance, the kinetic energy is 
measured in units of $m{\tilde\lambda}^2/\tau_0^2$. 
All simulations in I were carried with $n=c=0.3676$
and $\tilde\lambda=1$, $m=1$ and $\tau_0=1$. With
these conventions, the comparison with the molecular
dynamics uses $a_{WS}=0.866$, $C_2\alpha_s=0.216$
and $\Gamma=0.25/T$.

\vskip 1.0cm

{\bf Acknowledgments.\,\,}

This work was partially supported by the US-DOE grants DE-FG02-88ER40388
and DE-FG03-97ER4014.

%%%%%%%%%%%%%%%%%%%%%%%%%%%%%%%%%%%%%%%%%%%%%%%%%%%%%%%%%%%%%%%%%%%

\end{narrowtext}

\begin{thebibliography}{99}

\bibitem{Shu_QGP}
E.~V.~Shuryak, {Phys. Lett.} {\bf B78} (1978) 150;   
{Yadernaya Fizika } {\bf 28}  (1978) 796;
{Phys.Rep.} {\bf 61} (1980) 71. 

\bibitem{BP}E.~Braaten and R.~D.~Pisarski,
Nucl.\ Phys.\ B {\bf 337} (1990) 569.


\bibitem{hydro}
D.~Teaney, J.~Lauret and E.~V.~Shuryak,
Phys.\ Rev.\ Lett.\  {\bf 86} (2001) 4783;
P.F. Kolb, P.Huovinen, U. Heinz, H. Heiselberg,
{ Phys. Lett.} {\bf B500} (2001)  232;
Review in P.~F.~Kolb and U.~Heinz,
{\tt nucl-th/0305084}.


\bibitem{Shu_liquid}
 E.~Shuryak,  Prog.\ Part.\ Nucl.\ Phys.\  {\bf 53} (2004) 273.



\bibitem{SZ_newqgp}
E.~V.~Shuryak and I.~Zahed,
Phys.\ Rev.\ C {\bf 70} (2004) 021901;
Phys.\ Rev.\ D {\bf 70} (2004)  054507.

\bibitem{charmonium}
M. Asakawa and T. Hatsuda,
Nucl. Phys. {\bf A715} 863c (2003);
S.~Datta, F.~Karsch, P.~Petreczky and 
I.~Wetzorke, Nucl.\ Phys.\ Proc.\ Suppl.\  {\bf 119} (2003) 487.


\bibitem{GSZ_I} B.A. Gelman, E.V. Shuryak and I. Zahed,  
{\tt nucl-th/0601029}.

\bibitem{SOLUTES}
J.M. Caillol, {\tt cond-mat/0305465} and references therein.

\bibitem{DARBOUX} 
K. Johnson, Ann. Phys. (N.Y.) {\bf 192} (1989) 104.

\bibitem{Wong} 
S.K. Wong, Nuovo Cimento {\bf A 65} (1970) 689.

\bibitem{potentials} 
O.~Kaczmarek, S.~Ejiri, F.~Karsch, E.~Laermann and F.~Zantow,
{\tt hep-lat/0312015}.

\bibitem{Ichi} 
S.Ichimaru, H.Iyetomi and S.Tanaka, Phys.Rep.149 (1986) 92-205

\bibitem{solts} 
J.P.Hansen and I.R.McDonald, Phys.Rev.A11 (1975) 2111.

\bibitem{BRILLIANTOV}
N.~Brilliantov, {\tt cond-mat/9805358}

\bibitem{Shklovskii} A.Yu.Grosberg, T.T.Nguyen and B.I.Shklovkii,
cond.-mat/0105140.

\end{thebibliography}
\end{document}